\documentclass[twocolumn,aps,pra]{revtex4}

\usepackage{epsfig}
\usepackage[english]{babel}
\usepackage{latexsym}
\usepackage{graphics}
\usepackage{subfigure}
\usepackage{graphicx}
\usepackage{dcolumn}
\usepackage{amsmath}
\usepackage{hyperref}
\usepackage{amssymb}
\usepackage{color}

\begin{document}
\title{Characterizing Sub-Cycle Electron Dynamics of Polar Molecules by Asymmetry in Photoelectron Momentum Distributions}

\author{J. Y. Che$^{1}$, C. Chen$^{1}$, S. Wang$^{1,2,\dag}$, G. G. Xin$^{3}$, and Y. J. Chen$^{1,*}$}

\date{\today}

\begin{abstract}

Strong-field ionization of polar molecules contains rich dynamical processes such as tunneling, excitation, and Stark shift.
These processes occur on a sub-cycle time scale and are difficult to distinguish in ultrafast measurements.
Here, with a developed strong-field model considering effects of both Coulomb and permanent dipole,
we show that photoelectron momentum distributions (PMDs) in orthogonal two-color laser fields
can be utilized to resolve these processes with attosecond-scale resolution.
A feature quantity related to the asymmetry in PMDs is obtained,  with which
the complex electron dynamics of polar molecules in each half laser cycle is characterized and
the subtle time difference when electrons escaping from different sides of the polar molecule is identified.

\end{abstract}

\affiliation{1.College of Physics and Information Technology, Shaan'xi Normal University, Xi'an, China\\
2.School of Physics, Hebei Normal University, Shijiazhuang, China\\
3.School of Physics, Northwest University, Xi'an, China}
\maketitle

\section{Introduction}
Tunneling, one of the basic processes in quantum mechanism, is the dominant mechanism in strong-field ionization of atoms and molecules \cite{Keldysh,ADK}.
For polar molecules with a permanent dipole,  besides tunneling, strong-field ionization also contains other important processes,
such as excitation \cite{Bian} and Stark shift \cite{Etches}, greatly enriching strong-field physics \cite{Vos,Wustelt}.
Experimentally resolving these  processes is the first step for precisely measuring and controlling ultrafast electron motion of polar molecules.
Such measurements and controls are at the heart of  attosecond science and technology \cite{Krausz,Vrakking}.

On the whole, these processes occur on a sub-cycle time scale and are strongly coupled together,
making a quantitative identification of them through experimental observables such as photoelectron momentum distributions (PMDs) difficult \cite{Wang2020}.
In comparison with one-dimensional cases, two-dimensional laser fields have shown capability of higher resolution in ultrafast measurements \cite{Hasbani2}.
For example, PMDs in elliptical  \cite{Eckle,Hatsagortsyan,Undurti,Han} or orthogonal two-color (OTC)  \cite{Kitzler,Shafir,Zhang,Xie,Milo} laser fields,
can be utilized to  timing the ionization process
of atoms and symmetric molecules with attosecond resolution.
In these measurements, applicable strong-field models, which are capable of distilling dynamical information from PMDs, play a key role.

Here, we study strong-field ionization of polar molecules through numerical solution of time-dependent Schr\"{o}dinger equation (TDSE).
We show that with a developed strong-field model, which considers effects of both Coulomb potential and permanent dipole, 
OTC laser fields can be used to probe sub-cycle electron dynamics of polar molecules in ionization.
For  PMDs in different quadrants, comparisons of TDSE results with  model predictions
allow us to resolve not only tunneling but also excitation and Stark shift in ionization with high time resolution.
In particular, a feature quantity, which is associated with the ratio of PMDs of different quadrants, is obtained.
It  characterizes contributions of different processes to ionization in a half laser cycle, with providing a feasible tool for attosecond probing of polar molecules.

We begin our discussions with the simple case of HeH$^+$, then we validate our results for more complex cases such as CO and BF.

The TDSE of model polar molecules, including HeH$^+$ in Born-Oppenheimer (BO) and non-BO cases, CO and BF, etc., is solved with the spectral method \cite{Feit}.
Relevant numerical details including model potentials and grid sizes used in simulations are introduced  in \cite{Wang2017}.
Analytically, we use a modified model which arises from strong-field approximation (SFA) and electron-trajectory theory  \cite{Lewenstein2,Becker2002}
and includes effects of both Coulomb potential \cite{MishaY,Goreslavski,yantm2010} and permanent dipole (PD) \cite{Dimitrovski}.
For convenience, we call this modified model MSFA-PD. For comparison, we also use a simplified version which neglects the Coulomb effect in MSFA-PD and we call this version SFA-PD.
The details for these strong-field models associated with polar molecules can be found in \cite{Wang2020}.

The OTC electric field  $\mathbf{E}(t)$  used here consists of a strong fundamental field $E_{x}(t)$ and a weak second-harmonic field $E_{y}(t)$,
with \cite{Gao2017}
$\mathbf{E}(t)=\vec{\mathbf{e}}_{x}E_{x}(t)+\vec{\mathbf{e}}_{y}E_{y}(t)$,  $E_{x}(t)=f(t)E_{0}\sin(\omega_{0}t)$ and
$E_{y}(t)=\eta f(t)E_{0}\sin(2\omega_{0}t+\pi/2)$. $\vec{\mathbf{e}}_{x}$($\vec{\mathbf{e}}_{y}$) is the unit vector along the $x(y)$ axis.
$E_{0}$ is the laser amplitude relating to peak intensity \textit{I} of  $E_{x}(t)$.
$\eta$ is the amplitude ratio of $E_{y}(t)$ to $E_{x}(t)$.
$\omega_{0}$ is the laser frequency of $E_{x}(t)$ and $f(t)$ is the envelope function.
Here, the value of $\eta=0.5$ is used,  implying that the component $E_{x}(t)$ plays a dominant role in ionization.
We assume that the molecular axis is oriented parallel to $\vec{\mathbf{e}}_{x}$ and the heavier (lighter) nucleus is located on the right (left) side.
We use trapezoidally shaped laser pulses with a total duration of fifteen cycles of $E_{x}(t)$, which are linearly turned on and off
for three optical cycles, and then kept at a constant intensity for nine additional cycles.

\begin{figure}[t]
\begin{center}
\rotatebox{0}{\resizebox *{8.5cm}{8cm} {\includegraphics {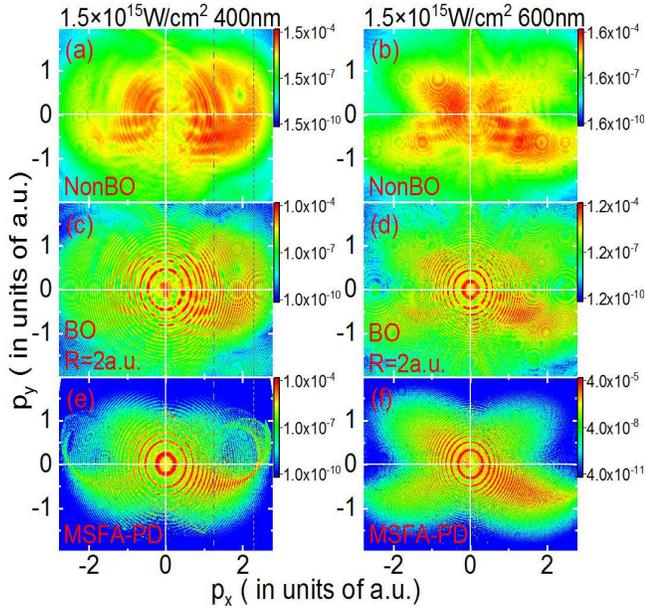}}}
\end{center}
\caption{PMDs of HeH$^{+}$ in OTC for $\lambda=400$ nm (the left column) and $\lambda=600$ nm (right) of the fundamental field,
obtained with different methods:
(a) and (b): TDSE-nonBO; (c) and (d): TDSE-BO with  $R=2$ a.u.;
(e) and (f): MSFA-PD with $R=2$ a.u..
The vertical lines are used to guide the eyes.
The laser intensity used is $I=1.5\times10^{15}$W/cm$^{2}$. The $\log_{10}$ scale is used here.}
\label{fig:g1}
\end{figure}

\section{Cases of HeH$^+$}
In Fig. \ref{fig:g1}, we present PMDs of HeH$^{+}$ in OTC fields obtained with different methods.
We focus on two laser wavelengths of $E_{x}(t)$, the short one of $\lambda=400$ nm (the left column) and the long one of $\lambda=600$ nm (right).
The TDSE results for  HeH$^{+}$ in non-BO cases are plotted in the first row of Fig. 1.
It can be observed that  for  $\lambda=400$ nm in Fig. 1(a), results in  all of the first (Q1), the second  (Q2) and the fourth (Q4) quadrants show large amplitudes.
By comparison, for  $\lambda=600$ nm in Fig. 1(b), results only  in Q2 and Q4  show large amplitudes.
It has been shown that vibrating HeH$^+$ stretches rapidly towards larger internuclear distances $R$ due to vibration excitation \cite{Wustelt,Liwy,Yue}.
For the present cases, with calculating R-dependent ionization probability as in  \cite{Liwy},
our analyses show that the main contributions to ionization occur around  $R=2$ a.u. ($I_p=1.44$ a.u.), somewhat larger than the equilibrium separation of $R=1.4$ a.u. ($I_p=1.66$ a.u.).
In the second row of Fig. 1, we show results of HeH$^+$ in BO cases with $R=2$ a.u..
The BO results with striking asymmetry are similar to the corresponding non-BO ones.
In the following discussions, for simplicity, we concentrate on BO results.

In the third row of Fig. 1, we show the results of MSFA-PD, relating to tunneling and PD-induced Stark shift, at the distance of $R=2$ a.u..
At this distance, the value of PD \cite{Wang2020} calculated is $D=-0.36$ a.u..
The model predictions match the corresponding TDSE results in most cases. For example, results in Fig. 1(c) show a bright tail which  sweeps from Q4 to Q1.
This characteristic is reproduced in Fig. 1(e). This tail is absent in both Figs. 1(d) and 1(f). In addition, in Figs. 1(e) and 1(f),
the MSFA-PD predicts large amplitudes for Q2 and Q4, in agreement with TDSE results in Figs. 1(c) and 1(d).
The remaining difference between TDSE and MSFA-PD is that results in Q1 in Fig. 1(c) show relatively larger amplitudes than those in  Fig. 1(e).
We will return to this point later.

\begin{figure}[t]
\begin{center}
\rotatebox{0}{\resizebox *{8.5cm}{6cm} {\includegraphics {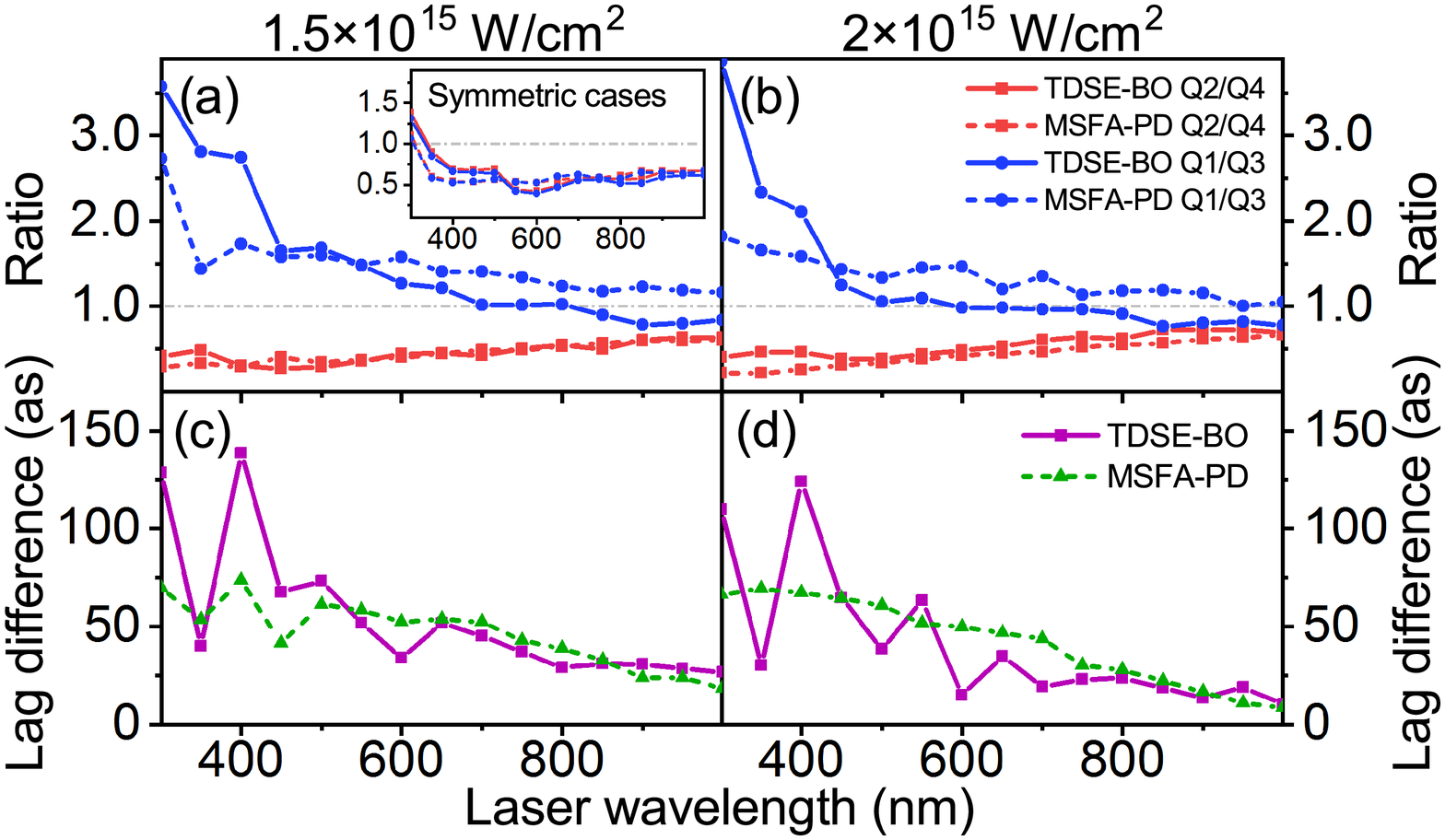}}}
\end{center}
\caption{
Ratios of Q1/Q3 and Q2/Q4 (a,b)   for PMDs of  HeH$^+$ in OTC and
the differences of the Coulomb-induced ionization time lag between the first and the second half laser cycles (c,d)
as functions of the fundamental wavelength,
at $I=1.5\times10^{15}$W/cm$^{2}$ (a,c) and $I=2\times10^{15}$W/cm$^{2}$ (b,d), obtained with TDSE-BO and MSFA-PD at $R=2$ a.u..
The inset in (a) shows these two ratios for a symmetric model molecule with  similar $I_p$ to HeH$^+$.
}
\label{fig:g3}
\end{figure}

Deeper insights are obtained when we further calculate the ratio of the total amplitudes in Q1 to those in Q3 (Q1/Q3) and that of Q2/Q4.
In the first row of Fig. 2, we show  these ratios of HeH$^+$ at $R=2$ a.u., obtained with TDSE-BO and MSFA-PD for different laser parameters.
First, both TDSE and model predictions in Figs. 2(a) and 2(b) of different laser intensities show that ratios of Q1/Q3 are remarkably larger than those of Q2/Q4, especially for short laser wavelengths.
Secondly,  results of MSFA-PD are very near to the TDSE ones for Q2/Q4, and they differ somewhat from each other for cases of Q1/Q3.
This difference is more remarkable when the laser wavelength is shorter, in agreement with the phenomenon discussed in Figs. 1(c) and 1(e).
By comparison, for a symmetric molecule with ${D}\equiv0$, these two ratios of Q1/Q3 and Q2/Q4 calculated with TDSE and MSFA-PD are similar to each other, as shown in the inset in Fig. 1(a).
We therefore  conclude that the difference between ratios of Q1/Q3 and Q2/Q4 for HeH$^+$ is closely associated with the PD effect.
In addition, as the MSFA-PD neglects the excitation, we anticipate that the difference between TDSE and MSFA-PD for ratios of Q1/Q3 at short laser
wavelengths discussed in Figs. 2(a) and 2(b) arises from the excited-state effect.

Next, we perform analyses to understand the above results.
First of all, we give a simple introduction on the PD effect \cite{Wang2020}. In the first row of Fig. 3, a sketch of laser-dressed bound states related to the PD effect of HeH$^+$ in a laser cycle is presented.
For the first half laser cycle with $E_x(t)>0$ (such as 6T to 6.5T, $T=2\pi/\omega_0$),
the laser polarization of the fundamental field of OTC is antiparallel to the PD,
and the asymmetric potential is bent along the H side.
In this case, the field-free ground state $|0\rangle$ is dressed up and
the first excited state $|1\rangle$ is dressed down, making ionization from the ground state (the ground-state ionization channel) easier to occur, as seen in Fig. 3(a).
It is worth noting that for the antiparallel case, the electron located in the dressed ground state $|0'\rangle$ is also easier to be pumped into the dressed excited state $|1'\rangle$.
The excited electron can survive the falling part of the laser field (as indicated by the vertical arrow in Fig. 3(g))
and ionize from the excited state in the following half laser cycle, opening the excited-state ionization channel \cite{Wang2017}.
One can expect that for shorter laser wavelengths, the excitation is easier to occur and the contributions of the excited-state channel to ionization are also more remarkable.

The situation reverses for the following half laser cycle with $E_x(t)<0$ (such as 6.5T to 7T),
where the ground state is dressed down and the excited state is dressed up, making both ionization and excitation from the ground state  more difficult to occur, as shown in Fig. 3(b).
These different effects of PD in the first and the following half  cycles lead to that the ionization dynamics of HeH$^+$ differ remarkably in one laser cycle.
One of the main aims of the paper is to resolve the complex sub-cycle ionization dynamics of HeH$^+$ from PMDs.

\begin{figure}[t]
\begin{center}
\rotatebox{0}{\resizebox *{8.5cm}{8cm} {\includegraphics {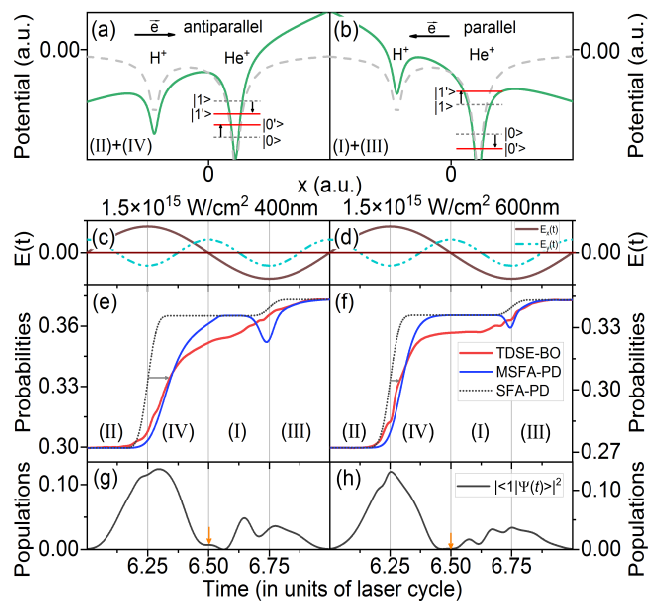}}}
\end{center}
\caption{
Sketch of the PD effect (the first row) and the OTC electric field $\mathbf{E}(t)$  (the second row) in one laser cycle of the fundamental field.
In the third row, we show that the comparison of time-dependent ionization probabilities of HeH$^+$ at $R=2$ a.u., calculated with TDSE-BO,
MSFA-PD and SFA-PD for different laser parameters, as shown. Here,  the model
curves are multiplied by a vertical scaling factor to match the TDSE-BO one.
The corresponding populations of the first excited state are plotted
in the fourth row.
The time region is divided into four parts of I to IV, relating to PMDs in Q1 to Q4.
When the laser polarization is antiparallel (a) or parallel (b) to the PD directing from the He nucleus to the H nucleus,
the energy of the two lowest electronic states $|0\rangle$ and $|1\rangle$ of HeH$^+$ is dressed differently. See the context for details.
}
\label{fig:g2}
\end{figure}

In the following, we discuss the ionization mechanism of HeH$^+$
with analyzing the time-dependent ionization probability (continuum population) $P(t)$ in a laser cycle.
We focus on the TDSE-BO cases in Fig. 1.

For comparison,  in the second row of Fig. 3, we show the OTC electric fields $E_x(t)$ and $E_y(t)$, with
dividing the time region of one laser cycle of $E_x(t)$ into four parts of I to IV. From a semiclassical view of point \cite{Corkum},
electrons which ionize in the time region of I to IV, will accordingly contribute to PMDs in Q1 to Q4 \cite{Xie}.
The TDSE probability $P(t)$ is obtained with excluding the components of  bound states of the field-free Hamiltonian $H_0$ from the whole wave function $|\Psi(t)\rangle$ at each instant $t$.
The model one  is obtained with evaluating the whole weights of the electrons, the instantaneous energy of which is larger than zero at the time $t$ \cite{Xie}.
Relevant results are presented in the third row of Fig. 3.

For results in Figs. 3(e) and 3(f), 1) all of the curves show that the ionization is  strong (weak) in the first (second) half laser cycle.
This asymmetry ionization phenomenon in one laser cycle can be well understood with the PD effect discussed above.
It also suggests that the PMDs  have larger amplitudes in Q2 and Q4 (related to ionization in regions II and IV) than those in Q1 and Q3
(regions I and III).
2) The probability curves of SFA-PD show a remarkable increase around $t=6.25T$ and $t=6.75T$ at which the electric field $E_x(t)$ arrives at its peak.
By comparison, for the first half cycle, the TDSE and MSFA-PD curves show a striking increase around a time later than $t=6.25T$.
This phenomenon has been attributed to the Coulomb induced ionization time lag 
between the instants of electrons tunneling out of the laser-Coulomb-formed barrier and  being free \cite{Xie}.
On average, the value of this lag can be taken as the time difference between the maximum of the electric field and the instant around which the
ionization increases remarkably, as the horizontal arrows show.
This lag has important influences on strong-field dynamics of the tunneling electron \cite{Xie,Wang2020,Wangs2020}.
The lag phenomenon is not noticeable for the second half laser cycle. Due to this large lag in the first half cycle, the ionization is stronger in region IV  than in region II, suggesting that
PMDs have larger amplitudes in Q4 than in Q2.
3) A careful analysis also tells that the TDSE curve in Fig. 3(e) of $400$ nm also shows a small increase around $t=6.5T$.
This increase disappears in Fig. 3(f) of the long wavelength case of $600$ nm. It is also absent for the predictions of MSFA-PD.
This small increase arises from the contributions of the first excited state which has a population at $t=6.5T$ with $E_x(t)=0$
for the short-wavelength case, as shown in Fig. 3(g).
Due to the excited-state contributions, the ionization is stronger in region I than in region III for $\lambda=400$ nm, indicating that
PMDs in Q1 have larger amplitudes than in Q3.
For the long wavelength case of $\lambda=600$ nm, the excited state can not survive the falling part of the laser field, as Fig. 3(h) shows.
In addition, this lag in the second half laser cycle is  small, region I and region III have comparable contributions to ionization,
indicating that PMDs in Q1 and Q3 have similar amplitudes.

With the above analyses, we can understand the OTC results in Fig. 1 and Fig. 2 as follows.
1) Due to the asymmetry ionization related to the PD effect, PMDs of polar molecules usually have larger amplitudes in Q2 and Q4 than in Q1 and Q3.
2) Due to the large Coulomb-induced ionization time lag in the first half laser cycle (in which electrons exit the barrier along the H side), the ratio of Q2/Q4 is usually smaller than 1.
3) This lag is small in the second half laser cycle (exiting along the He side) and accordingly the ratio of Q1/Q3 is near 1.
4) For short laser wavelengths, the excited-state ionization channel plays an important role in ionization. This channel increases the amplitudes in Q1, making
the ratio of Q1/Q3 in short-wavelength cases remarkably larger than 1. With the increase of the wavelength, the excited-state effect decreases \cite{Wang2017} and
the Coulomb-induced time lag also decreases \cite{Xie}, making that these ratios of Q1/Q3 and Q2/Q4 become near to each other.
5) This large (small) time lag in the first (second) half laser cycle reflects  the energy of the ground state  dressed up (down).
Specifically, for the case  dressed up (down), the tunneling exit is nearer  to (farther away from) the nuclei.
Accordingly, the Coulomb effect at the exit position is stronger (weaker) and the Coulomb induced time lag after the electron
exits the barrier along the H (He) side is also larger (smaller).
Therefore, the ratios of Q2/Q4 vs Q1/Q3, related to  the attosecond time lag of H vs He sides
and  excited-state contributions,  characterize sub-cycle electron dynamics of polar molecules.
In Figs. 2(c) and 2(d), we show the lag difference between  H and He sides. This lag difference decreases (from about  100 to about 20 attoseconds) with
increasing the laser wavelength, in agreement with the trend of the difference  between these two ratios.
One therefore can resolve the lag difference through the ratio difference.

\begin{figure}[t]
\begin{center}
\rotatebox{0}{\resizebox *{8.5cm}{6cm} {\includegraphics {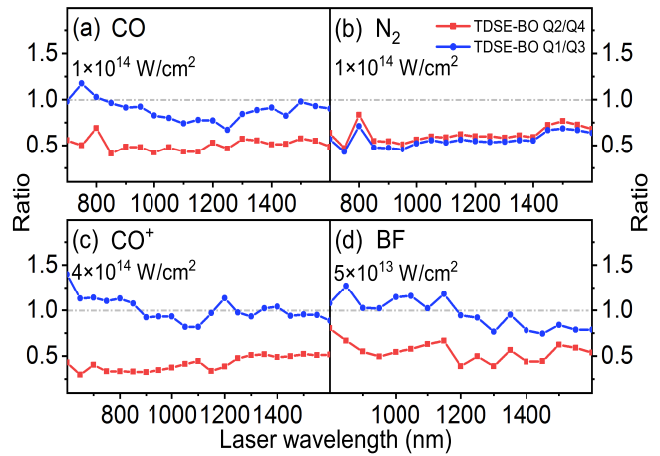}}}
\end{center}
\caption{
Ratios of  Q1/Q3 and Q2/Q4 as functions of the fundamental wavelength,
for PMDs of CO (a), N$_2$ (b), CO$^+$ (c) and BF (d) in OTC, obtained with TDSE-BO at the equilibrium separation.
The laser intensities $I$ used are as shown.
}
\label{fig:g4}
\end{figure}

\section{Other Cases}
To validate the above discussions, we have also performed simulations for molecules with more complex symmetries such as CO with $I_p=0.52$ a.u. and $R=2.135$ a.u.,
N$_2$ with $I_p=0.57$ a.u. and $R=2.079$ a.u., CO$^+$ with $I_p=0.98$ a.u. and $R=2.108$ a.u. and BF with $I_p=0.41$ a.u. and $R=2.385$ a.u..
In comparison with HeH$^+$, these molecules  can be  more easily manipulated in experiments.
As the molecules have heavier nuclei, TDSE-BO simulations at the equilibrium separation are applicable.
Considering that the molecules have different ionization potentials, we use different laser parameters
of the OTC fundamental field in our calculations.
Relevant results are presented in Fig. 4. Similar to results in Fig. 2,  for polar molecules of CO, CO$^+$ and BF, when the ratio  of Q2/Q4 is smaller than $1$,
the ratio of Q1/Q3 is near to 1. By comparison, for the symmetric case of N$_2$,  the ratios of Q1/Q3 and Q2/Q4 are near to each other and smaller than $1$.
These results support our previous discussions. In real experiments, perfect orientation is impossible \cite{Wang2020,Chen2013}. Our extended simulations show that
when the degree of orientation is larger than $40\%$, the main results in the paper such as the striking difference between behaviors of these ratios still hold.

\section{Conclusion}
In summary, we have studied the ionization of polar diatomic molecules with a large permanent dipole in strong OTC laser fields.
We have shown that in comparison with symmetric molecules,  ionization of  polar molecules  includes richer dynamical processes such as  tunneling, excitation and
Stark shift. In particular, due to the interaction of the permanent dipole and the laser field,
these processes differ remarkably when they occur in the first or the following half laser cycle,
with electrons escaping  from the side of the lighter or the heavier nucleus.
The subtle information of sub-cycle electron dynamics is well mapped in PMDs of  OTC fields.
With the help of a developed strong-field model which considers effects of both Coulomb potential and permanent dipole,
we are able to distill the information with high time resolution.

A feature quantity associated with ratios of PMDs in different quadrants is obtained,
which characterizes the dynamics difference of strong-field ionization in two consecutive half laser cycles and
can be used as a
tool in ultrafast measurements.
For example, it can be used to resolve the Coulomb-induced attosecond time lag (i.e., the delay between the instants of electrons exiting the barrier and being free),
which differs for exiting along different sides of the polar molecule
due to the effect of permanent dipole.
Our work opens new possibilities for probing and controlling ultrafast dynamics of polar molecules.

This work was supported by the National Natural Science Foundation of China (Grant No. 91750111),
and the National Key Research and Development Program of China (Grant No. 2018YFB0504400).

\end{document}